\documentclass{article}
\usepackage{spconf,amsmath,graphicx}

\usepackage{enumitem}
\setlist{nosep, leftmargin=14pt}

\usepackage{mwe} % to get dummy images
\usepackage{tabularx}
\usepackage{longtable}
\usepackage{afterpage}
\usepackage{multirow}
\usepackage{multicol}
\usepackage{float}
\usepackage{booktabs}

\begin{document}
\title{Region-based U-net for accelerated training and enhanced precision in deep brain segmentation}

%\name{Mengyu Li$^{1}$, Magnus Magnusson$^{1}$, Thilo van Eimeren$^{2,3}$, Lotta M. Ellingsen$^{*,1}$\thanks{*E-mail: lotta@hi.is}, \\
%and for the Alzheimer´s Disease Neuroimaging Initiative}

%\author{Mengyu Li$^{1}$, Magnus Magnusson$^{1}$, Thilo van Eimeren$^{2,3}$, Lotta M. Ellingsen$^{*,1}$\thanks{*E-mail: lotta@hi.is}, \\ and for the Alzheimer´s Disease Neuroimaging Initiative}

\name{ \begin{tabular}{c}
          Mengyu Li$^{1}$, Magnus Magnusson$^{1}$, Thilo van Eimeren$^{2,3}$, Lotta M. Ellingsen$^{*,1}$\thanks{*E-mail: lotta@hi.is}, \\
and for the Alzheimer´s Disease Neuroimaging Initiative
       \end{tabular} 
}
\address{$^{1}$ Faculty of Electrical and Computer Engineering, University of Iceland, Reykjavik, Iceland\\ $^{2}$ University of Cologne, Faculty of Medicine, Cologne, Germany\\ $^{3}$ University Hospital Cologne, Dept. of Nuclear Medicine and Dept. of Neurology, Cologne, Germany}

\maketitle

\begin{abstract}
Segmentation of brain structures on MRI is the primary step for further quantitative analysis of brain diseases. Manual segmentation is still considered the gold standard in terms of accuracy; however, such data is extremely time-consuming to generate. This paper presents a deep learning-based segmentation approach for 12 deep-brain structures, utilizing multiple region-based U-Nets. The brain is divided into three focal regions of interest that encompass the brainstem, the ventricular system, and the striatum. Next, three region-based U-nets are run in parallel to parcellate these larger structures into their respective four substructures. This approach not only greatly reduces the training and processing times but also significantly enhances the segmentation accuracy, compared to segmenting the entire MRI image at once. Our approach achieves remarkable accuracy with an average Dice Similarity Coefficient (DSC) of 0.901 and 95\% Hausdorff
Distance (HD95) of 1.155 mm. The method was compared with state-of-the-art segmentation approaches, demonstrating a high level of accuracy and robustness of the proposed method. 

\end{abstract}
\section{Introduction}
\label{sec:intro}
It is estimated that 5-8\% of the world population over the age of 60 (55 million) have dementia. This number is expected to rise to 78 million in 2030 and 139 million in 2050~\cite{who-dementia}. The boundaries between individual forms of dementia are often very indistinct, especially at early stages, making the accurate diagnosis of dementia a major challenge in clinical medicine with current diagnostic approaches. Certain dementias, however, show specific structural characteristics in the brain.

Parkinson-plus syndrome, also called atypical parkinsonism, is a group of neurodegenerative diseases that present with the classical features of Parkinson's disease (PD) but with additional features that distinguish them from PD. Parkinson-plus syndrome includes dementia with Lewy Bodies (DLB), progressive supranuclear palsy (PSP), and multiple system atrophy (MSA). DLB is reported as the second most common form of dementia after Alzheimer’s disease (AD)~\cite{Oppedal2019}. PSP and MSA are prototypical examples of diseases that often present with very specific structural characteristics in the brain. These include atrophy in deep-brain structures, such as the brainstem in PSP patients and the basal ganglia and the brainstem in MSA patients. Magnetic resonance imaging (MRI) plays an important role in distinguishing Parkinson-plus syndromes from PD and from each other. To accurately identify and differentiate between various causes of dementia, new diagnostic strategies involving imaging biomarkers and advanced analytical techniques may be necessary.

Various methods that perform whole-brain segmentation have been developed in recent years~\cite{brainsci11081055}. One such method is FreeSurfer~\cite{FISCHL2012774, iglesias2015bayesian}, which is widely used in MRI research and has been considered the benchmark for fully automated, quantitative analysis of brain anatomy from MRIs. Other approaches include multi-atlas segmentation methods~\cite{RN82, Wang2013, Huo2016}. More recently, fully automated brain segmentation approaches have been developed using deep convolutional neural networks (CNNs)~\cite{Mehta2017,huo2019whole,Shao2021,Atlason2022}. These methods have emerged as a major area of interest within the field of medical image analysis, providing both high accuracy and fast processing using a relatively small number of training images. Deep learning-based segmentation methods offer a significant speedup compared to both traditional segmentation methods and manual segmentation, reducing the processing time from hours to just a few minutes or even seconds.

One challenge of using deep learning-based methods in medical imaging is fitting the large data into the limited memory of the GPU. To achieve high accuracy with the deep learning models, a huge amount of MRI data is required for training, which can exceed the capacity of the GPU in addition to leading to excessive training time. A common solution is to use a patch-based approach, where the MRI images are divided into multiple patches. Patches can be loaded and processed one at a time, requiring much less memory. Patches can also be processed in parallel, allowing for efficient use of available computational resources, which can significantly speed up the segmentation process. However, patch-based methods can still take a considerable amount of training time, and the training time is closely related to GPU capacity. Huo et al.,~\cite{huo2019whole} reported a training time of 109 hours for a patch-based whole-brain three-dimensional U-net segmentation model trained on 5111 MRI scans~\cite{ronneberger2015, cicek2016}. The segmentation time for one scan was reported to be 15 minutes. 

Here we propose a multi-region-based, 3D CNN brain segmentation algorithm specific for deep-brain structures that have previously been associated with Parkinson-plus syndrome imaging biomarkers. A total of 83 T1-weighted MRI scans from three separate databases were used for training. The method provides a robust and fast segmentation of 12 deep-brain structures, including the brainstem substructures, i.e., the midbrain, pons, medulla oblongata, and superior cerebellar peduncle (SCP); the four ventricular compartments, i.e., the left and right lateral ventricles and the third and fourth ventricles; and four striatum structures, i.e., the left and right putamen and caudate nuclei. Compared to patch-based methods, this approach reduces both training and processing times and enhances precision in deep brain segmentation by enabling each U-net to be optimized around similar structures, such as cerebrospinal fluid (CSF) in the ventricular system and gray matter structures in the striatum. We hope that the method will contribute to identifying novel imaging biomarkers to aid in the early diagnosis of Parkinson-plus syndromes and other neurodegenerative diseases.

\section{METHODS AND MATERIALS}
\label{sec:format}

\textbf{Datasets:}
In this work, we used data from three separate data sources: 50 MRIs from Neuromorphometrics (NMM, www.neuromorphometrics.com), 73 MRIs from the Alzheimer’s Disease Neuroimaging Initiative (ADNI, adni.loni.usc.edu), and 26 MRIs from ASAP-CIR (Neuroimaging Study Group of the International Parkinson and Movement Disorder Society, www.movementdisorders.org). Each dataset was divided into training, validation, and test sets. The NMM dataset had 29 training, 6 validation, and 15 test subjects. The ADNI dataset had 44 training, 15 validation, and 14 test subjects. The ASAP-CIR dataset had 10 training, 5 validation, and 11 test subjects.

\textbf{Generation of training labels and ground truth data:}
While manually labeled data is considered the gold standard for brain structure segmentation and is frequently used as the ground truth for evaluating automated segmentation methods, the process of obtaining manually segmented images is extremely laborious and time-consuming. Deep learning-based segmentation methods often require a large amount of training data, making it impractical to have experts perform manual labeling for training. In this work, we used a multi-atlas segmentation method to automatically generate all of our training data~\cite{rudolph, magnus}. Furthermore, since manually labeled data was not available for ADNI and ASAP-CIR, we used a semi-manual segmentation approach to generate ground-truth data to be used for testing on these datasets, combining the expert's knowledge and visual intuition with the efficiency and consistency of an automated algorithm. The MRI scans and their corresponding labeled masks from NMM already contain 134 manually labeled anatomical structures, with the brainstem labeled as a single unit. For our segmentation purposes, we manually divided the brainstem into four substructures: midbrain, pons, medulla, and SCP based on the protocol in~\cite{iglesias2015bayesian}; Hence, fully manual labels were used as ground truth for NMM data.

\begin{figure*}[ht]  % Use 'figure*' to span both columns
  \centering
  \includegraphics[width=16cm]{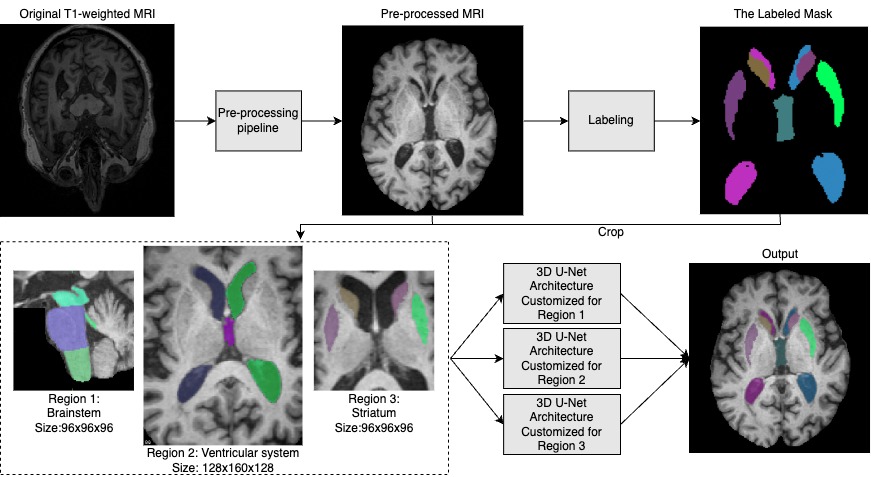}
  \caption{Workflow of the proposed region-based segmentation method.}
  \label{fig:workflow}
\end{figure*}

\textbf{The segmentation workflow and model architecture:}
Preprocessing is a fundamental step for precise segmentation using deep learning. We preprocessed all the raw MRIs by conducting the following steps: rigid registration to the MNI 152 atlas space,  N4 inhomogeneity correction~\cite{tustison2010n4itk}, and skull-stripping~\cite{skullstripping}. Since our training data were gathered from multiple different scanners and different imaging protocols, the image intensities had highly variable distribution values. To address this inconsistency between different scanners, the final pre-processing step was to normalize the intensities using a fuzzy c-means-based (FCM) normalization method~\cite{FCM}. Figure~\ref{fig:workflow} shows the workflow of our proposed method, including preprocessing, labeling, training, and prediction steps.

Our segmentation method is fully automated and implemented as a region-based approach. We divide the 12 aimed brain structures into three focal regions according to their anatomical position and relationships. Region 1 encompasses the brainstem, comprising the midbrain, pons, medulla, and SCP; Region 2 encompasses the ventricular system, comprising the left and right lateral ventricles, the 3rd ventricle, and the 4th ventricle; and Region 3 encompasses the striatum, comprising the left and right putamen and caudate nuclei. The MRI scans and the corresponding label masks were cropped from the original size of 240x285x240 voxels to 96x96x96 voxels for Region 1, 128x160x128 voxels for Region 2, and 96x96x96 voxels for Region 3. 

The architecture of each regional model is a modified 3D U-Net~\cite{ronneberger2015, cicek2016} and implemented in a similar fashion as our prior brainstem model~\cite{magnus}. Each model has layers of 32, 64, 128, 256, 512 channels and consists of five resolution steps. Each contracted step contains two 3x3x3 convolutional layers, ReLu activation, and a batch normalization layer after the latter convolutional layer. 
The model was pre-trained for 30 epochs using the Softmax-weighted cross-entropy loss and then trained for 200 epochs using the Dice loss function. The batch size of 4 was used for both pre-training and training. The learning rate was 0.01 using stochastic gradient descent and the momentum was 0.9.

\begin{figure*}[ht]
    \centering
    \includegraphics[width=15.6cm]{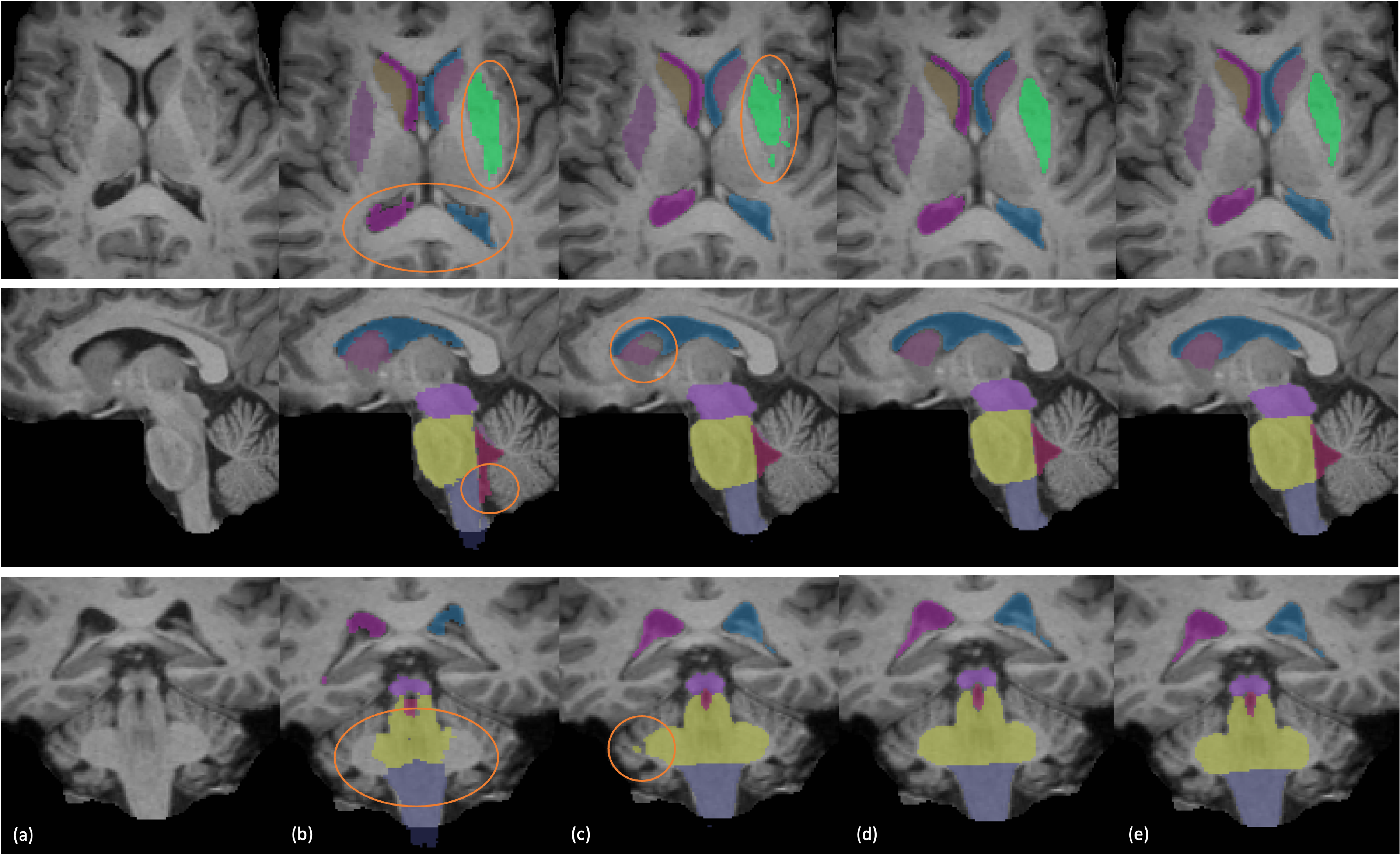}
    \caption{The T1-weighted MRI and its segmentation results from axial view (top), sagittal view (middle) and coronal view (bottom). From left to right are (a) T1-weighted MRI, (b) FreeSurfer v7 result, (c)patch-based method result, (d) region-based method result, (e) ground truth(semi-manual).}
    \label{fig:result}
\end{figure*}

\begin{table}[ht]
\caption{Mean DSC ± Standard Deviation (STD) over 40 test subjects, and across 12 structures using FreeSurfer, Patch-Based, and proposed Region-Based Segmentation Methods.}
\centering
\begin{tabular}{lccc}
\hline
 & \multicolumn{3}{c}{Method} \\
\cline{2-4}
&FreeSurfer & Patch-based & Region-based\\
\hline
Pons  & 0.809$\pm$0.279 & 0.950$\pm$0.011 & 0.969$\pm$0.007 \\
Midbrain & 0.762$\pm$0.258 & 0.904$\pm$0.018 & 0.936$\pm$0.016 \\
Medulla & 0.656$\pm$0.265 & 0.943$\pm$0.023 & 0.953$\pm$0.015 \\
SCP & 0.511$\pm$0.203 & 0.751$\pm$0.053 & 0.776$\pm$0.046 \\
R\_Ven & 0.831$\pm$0.198 & 0.921$\pm$0.051 & 0.924$\pm$0.055 \\
L\_Ven & 0.830$\pm$0.200 & 0.925$\pm$0.046 & 0.928$\pm$0.050 \\
T\_Ven & 0.725$\pm$0.215  & 0.836$\pm$0.078 & 0.862$\pm$0.109 \\
F\_Ven & 0.673$\pm$0.258 & 0.868$\pm$0.041 & 0.891$\pm$0.049 \\
R\_Cau & 0.754$\pm$0.201 & 0.802$\pm$0.129 & 0.880$\pm$0.046 \\
L\_Cau & 0.778$\pm$0.190 & 0.809$\pm$0.129 & 0.876$\pm$0.040 \\
R\_Pu & 0.800$\pm$0.189 & 0.870$\pm$0.093 & 0.909$\pm$0.044 \\
L\_Pu & 0.804$\pm$0.190 & 0.859$\pm$0.125 & 0.907$\pm$0.053 \\
\hline
Average & 0.744$\pm$0.220 & 0.870$\pm$0.067 & 0.901$\pm$0.044 \\
\hline
\end{tabular}
\label{tab:dsc}
\end{table}

\section{EXPERIMENTS AND RESULTS}
\label{sec:pagestyle}

\textbf{Region-based vs. patch-based approaches:}
Patch-based segmentation neural networks are commonly used in brain MRI segmentation to address GPU limitations. To compare with our region-based approach, we developed a patch-based CNN using the same training, validation, and test data as in our proposed region-based method, in addition to using the same training parameters for both models. For the development of the patch-based CNN, each 3D image was divided into 125 overlapping 3D patches, each patch with a size of 80x80x80 voxels, with a 50\% overlap of 40 voxels. Otherwise, the CNN architecture of the patch-based model was exactly the same as the proposed region-based method. 

We utilized two common metrics to quantitatively assess segmentation accuracy, i.e., DSC~\cite{dice1945measures} and HD95. We evaluated the mean DSC and HD95 over the 12 brain structures for both the patch-based method and our region-based method on our ground truth data (manual and semi-manual) from NMM, ADNI, and ASAP-CIR. The mean DSC of the patch-based method was 0.870$\pm$0.067, while our proposed region-based method achieved a mean DSC of 0.901$\pm$0.044, which is significantly higher than the DSC value from the patch-based approach ($p < 0.001$) according to a Wilcoxon signed-rank test. The mean HD95 of the patch-based method was 2.253 mm, whereas that of our proposed method was 1.155 mm. The smaller mean HD95 value indicates that the proposed method has closer correspondence between points in the segmented regions and their nearest neighbors in the ground truth, which is indicative of better segmentation accuracy.

Table~\ref{tab:dsc} shows the DSC for all 12 structures for both methods. For each structure, the region-based approach outperforms the patch-based method. Each region-based model's training was completed in approximately 4 hours on an Nvidia GeForce GTX GPU. In comparison, the patch-based method required approximately 7 days with the same training settings, yet, its segmentation accuracy was inferior to the region-based model.

\textbf{Comparison with state-of-the-art: }
FreeSurfer is the only (to our best knowledge) publicly available segmentation method that segments the 12 deep-brain structures at hand, including the 4 sub-structures of the brainstem. Given that FreeSurfer is considered a benchmark in brain segmentation, we computed the DSC for FreeSurfer  v7.3.2 based on our semi-manual ground-truth masks. Table~\ref{tab:dsc} shows the DSC for all 12 structures in FreeSurfer and our proposed region-based method. The mean DSC value over the 12 structures for FreeSurfer was 0.744$\pm$0.220, which is significantly lower than the DSC value from our approach ($p < 0.001$) according to a Wilcoxon signed-rank test. The FreeSurfer method had a mean HD95 of 3.207 mm, whereas our proposed method achieved 1.155 mm. In terms of both DSC and HD95, our proposed segmentation approach indicates a high level of accuracy and robustness. Our proposed method segments one subject in seconds compared to approximately 12 hours using FreeSurfer on the same machine. The segmentation results of state-of-art and our method are depicted in Figure~\ref{fig:result}.

\section{DISCUSSION AND CONCLUSION}
\label{sec:typestyle}
We presented a fully automated and robust region-based MRI segmentation method that speeds up the training process from days to hours for deep-learning-based, full-image segmentation approaches in addition to speeding up the segmentation process from hours to seconds compared to FreeSurfer and other multi-atlas segmentation approaches. Although neural network segmentation methods require long training times, once trained, they can segment a single MRI scan in just a few seconds. In contrast, the same segmentation process would take FreeSurfer several hours. Furthermore, the proposed method produces significantly more accurate results compared with the two competing methods shown in Table~\ref{tab:dsc}. The proposed method was specifically developed for automated parcellation of three focal regions comprising 12 deep-brain sub-structures, i.e., four sub-structures of the brainstem, the four ventricular compartments, the left and right putamen, and the left and right caudate. Finally, in addition to reducing training and processing times, the proposed method offers significant benefits, including the flexibility to customize and optimize the neural network architecture to suit specific brain regions with ease. Future work includes evaluating the proposed method on a larger cohort of healthy and Parkinson-plus syndrome patient data. It is our hope that the method can provide novel insights in addressing the important clinical challenge of early diagnosis of dementia and other neurodegenerative diseases.

\section{Acknowledgments}
\label{sec:acknowledgments}
This project was supported by RANNIS (The Icelandic Centre for Research) under grant 200101-5601.

\section{Compliance with ethical standards}
\label{sec:ethics}
This research study was conducted retrospectively using human subject data made available by NMM, ADNI, and ASAP-CIR. Ethical approval was not required as confirmed by the license attached with this data.

% References should be produced using the bibtex program from suitable
% BiBTeX files (here: strings, refs, manuals). The IEEEbib.bst bibliography
% style file from IEEE produces unsorted bibliography list.
% ------------------------------------------------------------------------- 
\bibliographystyle{IEEEbib}
\bibliography{strings,refs}

\end{document}